\documentclass[conference,10pt]{IEEEtran}
\usepackage{epsfig,epsf,rotating,setspace,latexsym,amsmath,amssymb,amsfonts,bm,theorem,subfigure,epstopdf}
\usepackage{cite,authblk}
\usepackage{bbm}
\usepackage{algorithm}
\usepackage[noend]{algpseudocode}
\usepackage{color}
\usepackage{mathtools}
\usepackage{soul}
\usepackage{hyperref}
\algrenewcommand\algorithmicforall{\textbf{foreach}}
\algrenewcommand\algorithmicindent{.8em}

\newtheorem{theorem}{Theorem}

\newtheorem{remark}{Remark}

\newenvironment{Proof}[1]{\medskip\par\noindent{\bf Proof:\,}\,#1}{{\mbox{\,$\blacksquare$}\par}}

\IEEEoverridecommandlockouts
\allowdisplaybreaks

\begin{document}
 
\title{Distributed Mixture-of-Agents for Edge Inference with Large Language Models}
 
\author[1]{Purbesh Mitra}
\author[2]{Priyanka Kaswan}
\author[1]{Sennur Ulukus}

\affil[1]{\normalsize University of Maryland, College Park, MD, USA}
\affil[2]{\normalsize Princeton University, Princeton, NJ, USA}

\maketitle

\begin{abstract}
Mixture-of-Agents (MoA) has recently been proposed as a method to enhance performance of large language models (LLMs), enabling multiple individual LLMs to work together for collaborative inference. This collaborative approach results in improved responses to user prompts compared to relying on a single LLM. In this paper, we consider such an MoA architecture in a distributed setting, where LLMs operate on individual edge devices, each uniquely associated with a user and equipped with its own distributed computing power. These devices exchange information using decentralized gossip algorithms, allowing different device nodes to \emph{talk} without the supervision of a centralized server. In the considered setup, different users have their own LLM models to address user prompts. Additionally, the devices gossip either their own user-specific prompts or augmented prompts to generate more refined answers to certain queries. User prompts are temporarily stored in the device queues when their corresponding LLMs are busy. Given the memory limitations of edge devices, it is crucial to ensure that the average queue sizes in the system remain bounded. In this paper, we address this by theoretically calculating the queuing stability conditions for the device queues under reasonable assumptions, which we validate experimentally as well. Further, we demonstrate through experiments, leveraging open-source LLMs for the implementation of distributed MoA, that certain MoA configurations produce higher-quality responses compared to others, as evaluated on AlpacaEval 2.0 benchmark. The implementation is available at: \url{https://github.com/purbeshmitra/distributed_moa}.
\end{abstract}

\section{Introduction}\label{section: introduction}
Large language models (LLMs) have significantly advanced the field of natural language processing through their ability to generate contextually relevant responses to general user queries~\cite{brown2020language, wei2022emergent, zhao2023survey, dong-etal-2024-survey, minaee2024large}. To further enhance the capabilities of LLMs, mixture-of-agents (MoA) has been proposed in \cite{wang2024mixture} as a method for collaborative inference with LLMs to produce high-quality responses to prompts. The inference is facilitated by the semantic exchange of information between neighboring LLM models, where different smaller open sourced models \emph{talk} with one another in natural language to produce high quality answers that rival cutting-edge proprietary models, such as OpenAI's GPT-4o, across various benchmarks. MoA improves upon the performance of individual models by employing a structure consisting of multiple proposer models and an aggregator model. The proposer LLMs propose new plausible responses for a given prompt, while the aggregator LLM either chooses the most appropriate answer or refines them by correcting the mistakes of the proposers. The effectiveness of this approach was demonstrated in \cite{wang2024mixture}, where the authors implemented a three-layer MoA system with three proposer LLMs in each layer. By utilizing a fixed prompt template for aggregation, which we refer to as the system prompt, it was shown that this method successfully generates high-quality responses comparable to GPT-4o on benchmarks such as AlpacaEval 2.0, Flusk 2.0, and MT-bench. In a subsequent work~\cite{li2024smoa}, sparse mixture-of-agents (SMoA) framework was proposed to further improve the efficiency, diversity, and scalability of multi-agent LLMs by introducing sparsity in agent interactions through mechanisms like response selection and early stopping.

\begin{figure}[t]
    \centerline{\includegraphics[scale=0.8]{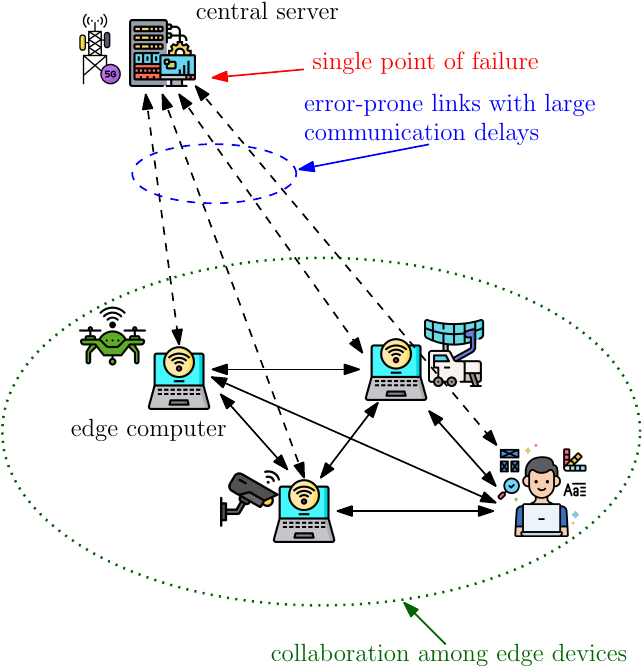}}
    \caption{An example of decision-making with edge devices. The central server with large compute power is directly connected to the edge devices. However, it is spatially separated from the edge devices, incurring large communication delays. Further, it is also a single point of failure, which can go offline due to link failures or adversarial attacks. The collaborations among edge devices, however, provides more diverse connections to more compute power than individual edge devices, thus making the system more robust.}
    \label{fig: edge_collab}
\end{figure}

\begin{figure}[t]
    \centerline{\includegraphics[scale=0.7]{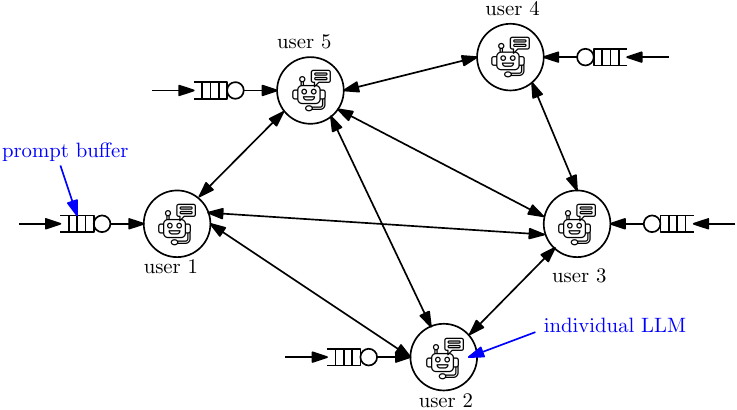}}
    \caption{System model of distributed MoA. Each device has their local prompts and which they infer by their local LLMs. Simultaneously, these prompts are sent to a few of their neighboring LLMs for inference and their responses are then aggregated by the original LLM. This response generation and aggregation process can continue for multiple rounds, constituting multiple layers of the MoA.}
    \label{fig: DMoA}
\end{figure}

An interesting observation about the MoA framework is that its proposer-aggregator format can be interpreted as different LLMs talking to one another using the semantics of natural language and refining the outputs to generate an improved response to a prompt. In the traditional communication setting, the main objective is accurate decoding of a data stream from a source on the bit-level, using tools such as error correcting codes. This approach is completely agnostic of the semantics of the communicated data, i.e., it does not take into account the accuracy of transferred information, like logical soundness, or factuality. In semantic communication~\cite{luo2022semantic, qin2021semantic, kalita2024large}, however, the goal is to efficiently relay the most accurate information content, where the bit-level representation at the receiver can be completely different from that of the source. This leads to an intuitive understanding of device-to-device communication on a semantic level, where the information can be corrected or refined by leveraging machine intelligence. Indeed, such semantic communication between agents is the key to achieving the hyper-connectivity in the upcoming 6G standards~\cite{lee23_6G, yang2022semantic}. With continuing progress in machine intelligence, we anticipate the deployment of more such agents in distributed edge devices, which will serve the user queries and often collaborate with other agents for accurate inference~\cite{wang2023adapting, li2024personal}. Such distributed systems enable users to utilize their own data and computational resources without relying on a centralized server, which is often a single point of failure susceptible to issues such as link disruptions and adversarial attacks, as well as incurring large communication delays with the edge devices. On the other hand, maintaining a diverse network of connections among edge devices enhances the overall computational robustness of the system, as shown in Fig.~\ref{fig: edge_collab}. This motivates us to explore the idea of \emph{semantic gossiping} within a distributed MoA setting. In our model, we consider multiple edge devices for computation or inference tasks, with each device hosting its own individual LLM. Thus, each of the agents or LLMs are spatially separated; see Fig.~\ref{fig: DMoA}.  

When multiple such devices are involved in semantic communication, their information exchange can be framed within the context of rumor-spreading or gossiping, where different devices communicate with one another to accomplish certain goal-oriented tasks without the supervision of any centralized server~\cite{demers87epidemic_gossip}. This process of information hopping between nodes incurs some latency within the system. There has been significant research focusing on the timing aspect of gossiping for a single source of fixed information~\cite{shah08monograph} and the timeliness of gossiping for a time-varying source~\cite{yates21gossip}. Subsequent works have explored how gossip network scaling impacts the timeliness of the overall system~\cite{kaswan2023age_review}. However, all these works only consider the latency aspect of the system, without taking the content or semantics of the data into account. To the best of our knowledge, there has not been any work that has considered both the semantics of the data being gossiped and its timeliness. This gap motivates us to formulate the goal of our paper of analyzing the capacity or stability of gossiping in the context of a distributed MoA system.

\begin{figure}[t]
    \centerline{\includegraphics[scale=0.58]{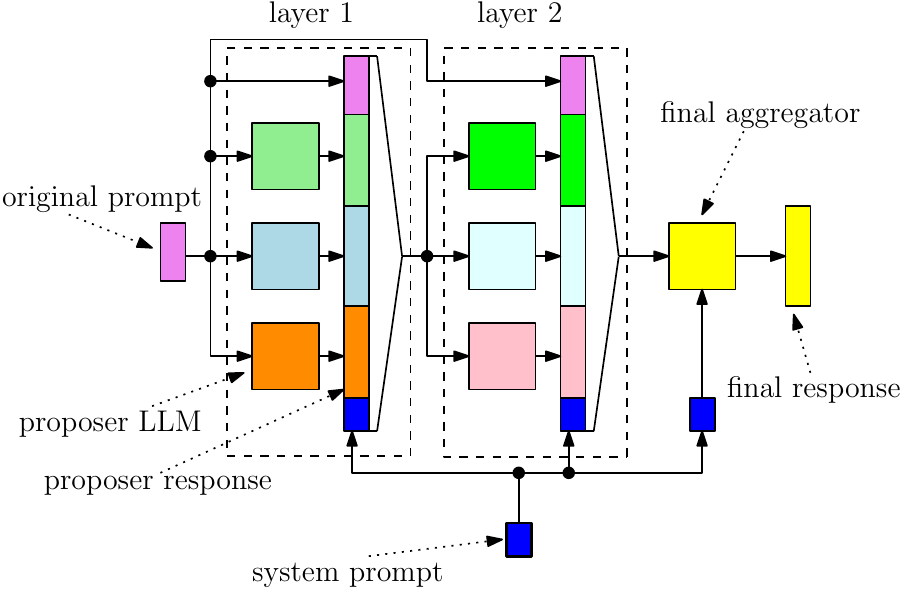}}
    \caption{An illustration of a 2-layer mixture-of-agents system. The original prompt is inferred via 3 proposer LLMs in each layer and is finally aggregated at a single aggregator LLM. From second layer onwards, the LLMs use the system prompt to generate refined outputs from the concatenated prompts.}
    \label{fig: MoA_model}
\end{figure}

This goal of this paper is to implement a distributed MoA network and analyze its queuing stability. In other words, we aim to achieve maximum possible accuracy while ensuring that the overall system latency remains bounded. This approach allows edge devices to participate in the MoA framework while operating under memory constraints. Since system performance is influenced by the prompt generation rate at different users, we analyze the queuing stability of the LLMs at various users under some reasonable assumptions. In particular, we assume that the prompts are generated for each user with a mean inter-arrival time, and these user LLMs can gossip their prompts for further assistance in the MoA format. The LLMs also have some mean prompt-inference time. We calculate the queuing stability as a function of this prompt generation rate and analytically determine whether the system is stable or if the queue size can grow exponentially. Additionally, we demonstrate through simulations how accuracy varies when the MoA structure is modified, such as by increasing the number of layers or the number of proposers within a layer.

\begin{figure*}[t]
    \centerline{\includegraphics[scale=0.7]{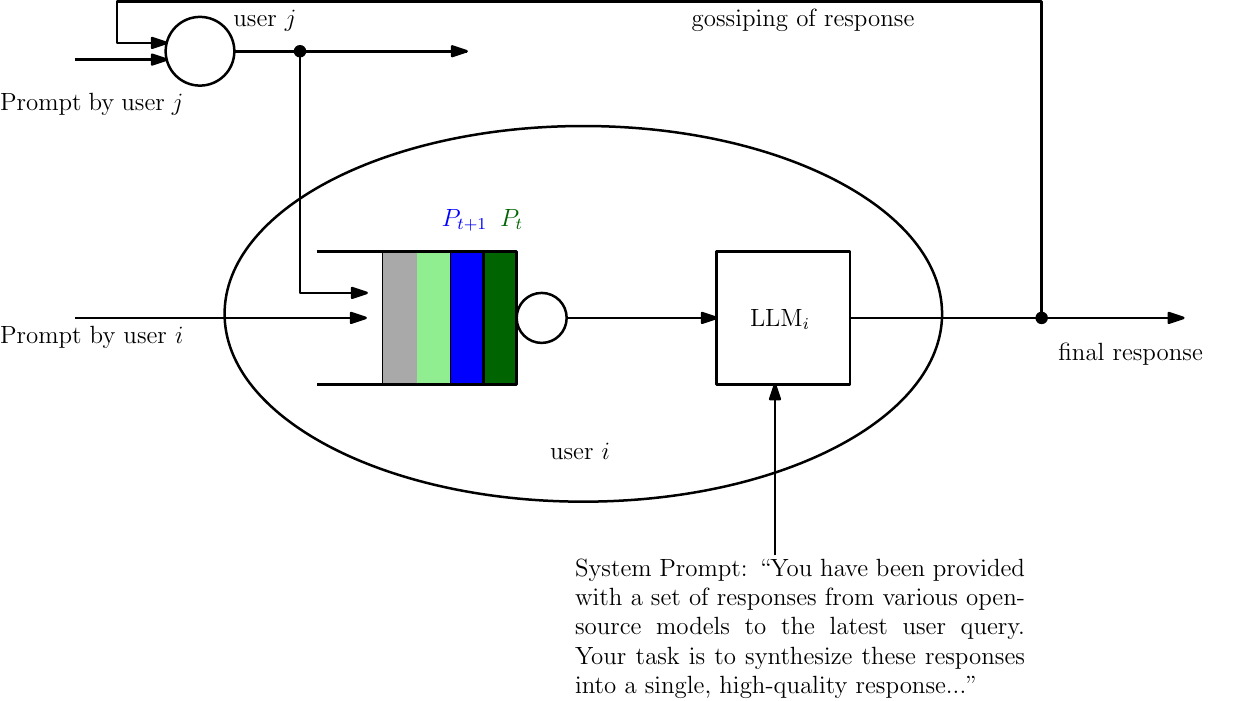}}
    \caption{An overview of user $i$ in distributed MoA. It has the LLM$_i$ in it, which sequentially infers prompts from the adjacent queue. The prompts are generated by user $i$ itself and also from its neighbour $j$. The output of the LLM is sent to the neighboring node for further assistance in inferring. The final output comes from aggregating over different responses at the LLM$_i$.}
    \label{fig: DMoA_network}
\end{figure*}

\section{System Model}\label{section: system model}
In this section, we describe the system model for the distributed mixture-of-agents (MoA). We consider a network of $n$ users, as illustrated in Fig.~\ref{fig: DMoA}, where each user has its own individual LLM at its edge device. These LLMs can exchange prompts with one another to infer collaboratively for improving the quality of prompt responses. The LLM $i$ gets prompts from its corresponding user $i$ at rate $\lambda$ as well as prompts from other LLMs for collaborative inference, which get stored in a first-come-first-served (FCFS) queue at the device. Once a prompt from the queue is inferred by an LLM, it is then either directly sent back to the user or stored in a buffer if further processing is needed; see Fig.~\ref{fig: DMoA_network}. For simplicity, we assume that the average LLM inference time is $\alpha$. We aim to analyze the conditions under which the queue remains stable as a function of $\lambda$ (the prompt generation rate) and $\alpha$ (the average inference time). Each user sends its prompts to the device queue for processing by its corresponding LLM, and also additionally selects, uniformly at random, $k$ out of other $n-1$ neighboring devices to transmit the prompts for inference. When the responses from the neighboring $k$ LLMs have been returned, the original user's LLM acts as an aggregator to synthesize a refined output. The aggregator LLM takes the original prompt and the collected responses as input, guided by the following system prompt for instruction: 

\emph{``You have been provided with a set of responses from various open-source models to the latest user query. Your task is to synthesize these responses into a single, high-quality response. It is crucial to critically evaluate the information provided in these responses, recognizing that some of it may be biased or incorrect. Your response should not simply replicate the given answers but should offer a refined, accurate, and comprehensive reply to the instruction. Ensure your response is well-structured, coherent, and adheres to the highest standards of accuracy and reliability. Do not add any additional comments about how you created these responses. Just synthesize these responses as instructed."}~\cite{wang2024mixture}. 

\begin{algorithm}[h]
  \caption{Distributed Mixture of Agents Algorithm}
  \label{algo: DMoA}
  \begin{algorithmic}[1]
    \State Prompt arrival rate is $\frac{\lambda}{n}$ at all the users.
    \For{$j\in[M]$}
        \For{$i\in[n]$}
            \Procedure{PromptArrival}{user $i$}
                \State Push the prompt into its LLM queue.
                \State If generated by user $i$, choose $k$ LLMs randomly to transmit.
            \EndProcedure
            \Procedure{ModelInference}{user $i$}
                \State Infer the generated prompt with expected time $\alpha.$
                \State If prompt is generated by user $j\neq i$, send the response back to $j$.
            \EndProcedure
            \Procedure{PromptConcatenation}{user $i$}
                \State Concatenate the responses from different LLMs for an inferred prompt.
                \State Push the concatenated prompt into the LLM inference queue.
                \State $j \leftarrow j+1$
                \State Send the concatenated prompt to $k$ randomly chosen
                LLMs if $j < M$. 
            \EndProcedure
        \EndFor
    \EndFor
  \end{algorithmic}
\end{algorithm}

\begin{table*}[t]
    \centering
    \caption{Accuracy, latency, and average queue for different MoA configurations with same LLM model with 0.7 temperature}
    \label{table: moa_single_prop}
    \begin{tabular}{|c|c|c|c|}
    \hline
    MoA configuration  &Accuracy (\%)  &Latency (seconds)  &Average queue size \\
    \hline
    $M=0, k=0$   &21.91  &94.01 &0.6 \\
    \hline
    $M=1, k=1$   &29.55  &329.92  &2.04  \\
    \hline
    $M=1, k=2$   &39.63  &484.81  &3.28 \\
    \hline
    $M=2, k=1$   &28.88  &603.43  &2.32 \\
    \hline
    $M=2, k=2$   &39.89  &790.01   &4.80 \\
    \hline
    $M=1, k=3$   &39.79  &578.88   &3.74 \\
    \hline
    $M=2, k=3$   &40.63  &1074.28  &5.78 \\
    \hline
    \end{tabular}
\end{table*}

This is essentially the basic structure of the MoA for a single layer. If multiple layers are employed, the prompt, along with the concatenated responses and system prompt, is again forwarded to $k$ randomly selected LLMs for further inference. The MoA process iterates through these layers until the final result is aggregated by the user’s LLM, as illustrated in Fig.~\ref{fig: MoA_model}. The complete distributed LLM communication protocol is shown in Algorithm~\ref{algo: DMoA}.

\section{Queuing Stability Calculation}\label{section: stability}
In this section, we analyze the queuing stability of the distributed MoA system. The following theorem establishes the conditions for system stability as a function of the prompt generation rate and the average inference time.

\begin{theorem}
The individual queues in the distributed MoA setting remain stable if the following condition is satisfied:
\begin{align}
    \alpha((k+1)M+1)\lambda < 1.
\end{align}
\end{theorem}

\begin{Proof}
In the distributed MoA setting, each LLM processes its own user's prompts that are arriving with rate $\lambda$. Additionally, it processes prompts from the other users with probability $\frac{\binom{n-2}{k-1}}{\binom{n-1}{k}} = \frac{k}{n-1}$. Hence, each LLM proposer in a single layer has the prompt arrival rate,
\begin{align}
    R^{prop}_{in}= \lambda + (n-1)\lambda \times \frac{k}{n-1} = (k+1)\lambda.
\end{align}
If there are $M$ number of layers in the MoA, then an LLM receives these prompts at a rate that is $M$ times the original rate. Thus, the input throughput for a device queue becomes,
\begin{align}
    R^{layer}_{in}= R^{prop}_{in}\times M = (k+1)M\lambda.
\end{align}
Additionally, for the aggregation step, the original user's LLM is used again, resulting in total contribution of $(k+1)M+1$ times the original throughput. Thus, the final input throughput generated from a single user can be expressed as,
\begin{align}
    R_{in}=((k+1)M+1)\lambda.
\end{align}
Based on our assumption that the average inference time of an LLM is $\alpha$, the output throughput of an LLM is $R_{out}=\frac{1}{\alpha}$. For a stable FCFS queue, the stability condition is known to be $R_{in} < R_{out}$. Consequently, the stability condition becomes,
\begin{align}\label{queue_stability}
    ((k+1)M+1)\lambda < \frac{1}{\alpha},
\end{align}
concluding the proof.
\end{Proof}

\begin{remark}
Note that a diverse set of LLMs can be used by different devices, and their average inference times will not necessarily be the same constant $\alpha$. Let us denote the average inference times as $\{\alpha_1, \alpha_2, \cdots, \alpha_n\}$ for all the $n$ LLMs. Then, the queuing stability condition depends on the inference rate of the slowest LLM, i.e., the LLM with the highest mean inference time. Hence, for a general case, \eqref{queue_stability} can be written as,
\begin{align}
    ((k+1)M+1)\lambda < \frac{1}{\alpha_{\max}}=\frac{1}{\max\{\alpha_1, \alpha_2, \cdots, \alpha_n\}}.
\end{align}
\end{remark}

\begin{table*}[t]
    \centering
    \caption{Accuracy, latency, and average queue for different MoA configurations with diverse LLMs}
    \label{table: moa_multi_prop}
    \begin{tabular}{|c|c|c|c|}
    \hline
    MoA configuration  &Accuracy (\%)  &Latency (seconds)  &Average queue size  \\
    \hline
    $M=0, k=0$   &23.62  &409.58  &1.71 \\
    \hline
    $M=1, k=1$   &39.41  &806.88  &2.36   \\
    \hline
    $M=1, k=2$   &39.52  &1112.33  &2.72  \\
    \hline
    $M=2, k=1$   &38.22  &1357.74  &1.83   \\
    \hline
    $M=2, k=2$   &40.51  &1971.50  &3.07   \\
    \hline
    $M=1, k=3$   &38.96  &1292.73  &2.92   \\
    \hline
    $M=2, k=3$   &55.55  &2274.59  &4.21   \\
    \hline
    \end{tabular}
\end{table*}

\section{Numerical Results}\label{section: numerical_results}
In this section, we present the experimental results with distributed MoA settings. We use four different open-source LLMs: Llama-3-70B, Qwen-1.5-72B, Mixtral-8x22B-v0.1 and dbrx-instruct, i.e., $n=4$. The prompt arrival rate for individual users is $\lambda=\frac{1}{4}$. We compute the accuracy in a subset of AlpacaEval 2.0 dataset.\footnote{Due to resource constraints, our evaluation focuses on 10 representative samples. This exploratory analysis provides preliminary insights into the distributed MoA’s capabilities, latency limitations, and trends, which align with the findings of the full simulation for MoA in \cite{wang2024mixture}.} The accuracy of the LLM responses are evaluated via GPT-4 API. Additionally, we evaluate the system's average queue size and average prompt latency. The implementation is available at: \url{https://github.com/purbeshmitra/distributed_moa}.

In our first experiment, we implement a system where all the users use a Qwen-1.5-72B model at their edge devices. All the models sample at their output with the temperature parameter $0.7$. This makes the output token streams different for different LLMs with the same prompt. We experiment with different number of layers $M$ and different number of proposers $k$ and find the accuracy in AlpacaEval 2.0 dataset and the corresponding average system latency. The results are summarized in Table~\ref{table: moa_single_prop}.

Additionally, we also calculate the average queue size, which measures the average number of prompts waiting in the queues for LLM inference. In Table~\ref{table: moa_single_prop}, we observe that there is a clear trade-off between the accuracy and the average queue-size. To obtain higher accuracy, it is necessary for the prompts to be processed with multiple LLMs, and thus the individual queue sizes must be larger to accommodate higher prompt traffic. As observed in the table, increasing the number of layers or proposers in a layer clearly deteriorates the timeliness performance of the system. The highest accuracy and latency are observed with a configuration of $M=2$ layers and $k=4$ proposers. Thus, for achieving higher accuracy, while keeping a constrained budget on processing time or user memory, requires some trade-off in the choice of parameters $M$ and $k$, as evident from the table.

Finally, we show the results for a distributed MoA with all four of the different LLMs combined. In this case, we use the four LLMs: Llama-3-70B, Qwen-1.5-72B, Mixtral-8x22B-v0.1 and dbrx-instruct for inferring the prompts. Each prompt is generated randomly at different users, and the users communicate with their neighbors for processing the prompt in a MoA format. We show the accuracy and latency for different model configurations, as well as the average queue size, for different values of $M$ and $k$ in Table~\ref{table: moa_multi_prop}. 

We observe that using diverse LLMs increases the accuracy. This is evident from the fact that keeping the number of proposers and the layers the same as in Table~\ref{table: moa_single_prop} increases the accuracy. We also observe similar trends of increasing accuracy with more number of layers and proposers per layer as in the single LLM case. Another observation is that the average queue size also increases with higher system latency, thus confirming the trade-off between accuracy and queue size. 

\section{Conclusion}\label{section: conclusion}
In this work, we implemented a distributed MoA framework for edge device collaboration with LLMs. Each edge device has their own individual LLM. If a device only uses its own LLM, its response accuracy remains poor. This accuracy can be increased if the LLMs collaborate together in a MoA framework. The MoA framework allows the LLMs to generate responses, which are further refined by other LLMs in multiple layers to generate higher quality responses. We showed how such a system can be implemented via gossiping prompts and responses in a decentralized manner. Our experimental simulations show that there exists a trade-off between accuracy and system latency. Furthermore, higher latency also necessitates the device queue size to be larger.

\bibliographystyle{unsrt}
\bibliography{reference}

\end{document}